\documentclass[debug]{rmaa}

\usepackage{paralist}
\usepackage{psfrag,color} 
\usepackage{graphicx}
\usepackage[latin1]{inputenc}
\usepackage{epstopdf}

     \title{AGB candidates in the field of $\gamma$ Cas.  \altaffilmark{1} }    
\author{
R. Nesci,       \altaffilmark{2} 
 T. Tuvikene, \altaffilmark{3}
C. Rossi,       \altaffilmark{4}
S. Gaudenzi,     \altaffilmark{2} 
 S. Galleti,      \altaffilmark{5} 
 P. Ochner,   \altaffilmark{6,7} 
 H. Enke   \altaffilmark{8} 
   }
   
\altaffiltext{1} {  Based on observations collected with  the Cassini Telescope at Loiano station of the INAF$-$Bologna Astronomical Observatory, the 67/90 Schmidt telescope and the Copernico Telescope of the INAF$-$Padova Astronomical Observatory, the 122-cm telescope of the Padova University.}
\altaffiltext{2}{  INAF/IAPS, Roma Italy.  }
\altaffiltext{3}{ Tartu Observatory, University of Tartu, Estonia.       }
\altaffiltext{4}{ INAF/ Osservatorio di Monte Porzio, Roma, Italy.          }
\altaffiltext{5}{ INAF-IASF,  Bologna, Italy.    }
\altaffiltext{6}{ INAF-Osservatorio Astrofisico Asiago, Italy.   }
\altaffiltext{7}{ Dipartimento di Fisica e Astronomia, Universita' di Padova, Italy.  }
\altaffiltext{8}{ Leibniz-Institut f\"ur Astrophysik, Potsdam (AIP), Germany.    }

\shortauthor{ Nesci \& al.}
\shorttitle{AGB candidates in Cassiopeia}

\fulladdresses{
\item Harry Enke:  Leibniz-Institut f\"ur Astrophysik, An der Sternwarte 16, 14482 Potsdam, Germany (henke@aip.de).
\item Silvia Galleti : INAF-IASF via Piero Gobetti 93/3, Bologna, Italy (silvia.galleti@oabo.inaf.it).
\item Silvia Gaudenzi and Roberto Nesci: INAF/IAPS, via Fosso del Cavaliere 100, 00133 Roma, Italy (silvia.gaudenzi@iaps.inaf.it); (roberto.nesci@iaps.inaf.it).
\item Paolo Ochner: INAF-Osservatorio Astrofisico, via Osservatorio Astronomico, 8, 36012 Asiago (Vi)
 and  Dipartimento di Fisica e Astronomia,  Universit\'a di Padova, via Marzolo, 8. I-35131 Padova, Italy
(paolo.ochner@oapd.inaf.it).
\item Corinne Rossi: INAF-Osservatorio Astronomico di Roma, Via Frascati 33, 00040, Monte Porzio Catone (RM), Italy (corinne.rossi@uniroma1.it).
\item Taavi Tuvikene:  Tartu Observatory, University of Tartu, Estonia  (taavi.tuvikene@to.ee).
  }

\listofauthors{ R. Nesci , T. Tuvikene  \& al.}
\indexauthor{Enke, H.}
\indexauthor{Galleti, S.}
\indexauthor{Gaudenzi, S.}
\indexauthor{Nesci, R.}
\indexauthor{Ochner, P.}
\indexauthor{Rossi, C.}
\indexauthor{Tuvikene, T.}
 
\abstract
{We report the spectroscopic and photometric monitoring of a sample of 530 candidate AGB stars in a 5 degrees field, selected from the IPHAS catalog: historic light curves were derived from Asiago IR plates taken in the years 1965 -1984. We found 10 Miras, 5 stars with long term trends, 3 semiregular and 3 irregular. 
Spectral types from CCD slit spectra gave 8 M-type, 7 C-type and 6 S-type stars. 
In the color-color plots made from IPHAS and 2MASS catalogs, the S-type and M-type stars occupy the same regions, while C-type stars are well separated. All C-type stars with  IR excess showed long term trends in their light curve.
Distances of the Mira stars, estimated from their periods and K magnitudes, gave a median value of 4.9 kpc with a large spread.  A comparison with astrometric parallaxes from Gaia DR2 is briefly discussed. 
}

\resumen{ 
En este artículo discutimos los resultados de un monitoreo  espectroscópico y fotométrico de una muestra de estrellas AGB  candidatas en un campo de 5 grados en el plano galáctico centrado en  gamma Cas.
Se seleccionaron 530 estrellas del catálogo IPHAS, con i $< 15$ ~y $ (r-i) >  1.7$ ~magnitud. Las curvas de luz históricas se derivaron de las placas  IR del Observatorio de Asiago tomadas en los años 1965-1984.
Solo 21 estrellas mostraron una variación significativa: 10 estrellas  fueron Miras regulares, 5 mostraron tendencias a largo plazo, 3  estrellas parecían SR y 3 estrellas irregulares.
Los tipos espectrales se derivaron de los espectros de rendija CCD  tomados en los observatorios de Loiano y Asiago: 8 estrellas son de  tipo M, 7 estrellas de tipo carbón y 6 estrellas de tipo S. Las  graficas de color-color se derivaron de los datos del catalogo IPHAS y  2MASS.
Todas las estrellas de carbono con un fuerte exceso de IR mostraron  tendencias a largo plazo en su curva de luz. Las estrellas tipo S y  tipo M ocupan las mismas regiones en los diagramas NIR de color y las  estrellas de carbono están bien separadas.
Las distancias de los Miras se estimaron a partir de sus períodos y  magnitudes K: se encontró una gran dispersión en la estimación de  distancias, con un valor medio de 4,9 kpc .
 }

\addkeyword{Stars: AGB}
\addkeyword{Stars: late-type}
\addkeyword{Stars: variables}

\begin{document}
\maketitle


\section{Introduction}  \label{sect:intro}
The AGB phase is the final stage of evolution before the star becomes a white dwarf after a {    brief} Planetary Nebula stage. Given their high infrared luminosity, AGB stars can be detected at large distances even on the galactic plane, where interstellar extinction is high, and serve as tracers of the stellar population. Moreover, if they are in the Mira phase, their intrinsic magnitude can be inferred from the observed period so that they can be used as distance estimators.

Substantial improvements on the accuracy of absolute luminosities
of a large number of stars {    is provided by the DR2 catalog from the Gaia mission, and future releases can possibly reach stars up to a distance of about 4 kpc.} Expanding the sample of Mira stars with well defined periods within such
distance is therefore very useful to improve their Period-Luminosity relation.


In this paper we report a variability study of very red stars selected in a 25 square degrees area centered on the bright star $\gamma$ Cas, at galactic coordinates b=$-$2$^o$, l=123$^o$: this area covers a part of the Perseus spiral arm of the Milky Way, at a distance of about 2.4 kpc from our solar system \citep{Rei14}. It was monitored by 87  plates of the Asiago Observatory taken with the 67/92/215 cm Schmidt telescope and  an emulsion+filter combination (Kodak I-N + RG5) which gives a passband  similar to the Cousins $I_\mathrm{C}$ filter. These plates were taken between 1967 and 1984, most of them before 1975 with rather uniform time sampling, while only a few were obtained in the last years. 

The field of $\gamma$ Cas was the fourth of a series of 4 fields observed by P. Maffei to search for Mira variables, the others being IC 1805 \citep{Gas91}, $\gamma$ Cyg \citep{Maf77}, and M16 \citep{Maf99} . 
The aim of the  work was a  systematic study of the  distribution  and the relative fractions of  the different types of late type variables in fields  lying at low galactic latitude but different  longitudes. {    This field was however not studied, so that no discoveries of variable stars  based on these plates were published.}

A study of these old plates has several advantages: first, they provide a continuous time coverage of about 7 years, corresponding to 7--8 cycles for a typical Mira star, large enough to define a robust period determination; second, it allows a comparison with modern observations over a couple of decades, enough to see if substantial variations in the light curves have occurred;  third, the used wavelength (800 nm) is less absorbed by interstellar matter and allows to explore a large volume in the plane of the Milky Way.
This dataset is therefore well suited for the study of Mira variables: {    a similar time span is covered by a recent study of Mira stars based on the ASAS project \citep{vogt16}, with objects south of $\delta$ +28$^o$.}

A trial test to explore the quality of these plates was made by \citet{Nes16} on a single known variable (V890 Cas) and was quite successful, motivating a larger effort with an automatic technique.

\section{Candidate selection}
\label{sect:sear}
The search for candidate AGB stars is most easily made by a color selection: For this purpose we used the IPHAS DR2 catalog \citep{Bar14}, which provides $r$, $i$ (Sloan filters) and $H_\alpha$ magnitudes down to $i\sim 20$ mag for a large part of the Milky Way: the zero point of these magnitudes is on the Vega scale.

To determine the proper $r-i$ color for candidate AGB stars a first guess is given by \citet{Sal09} who indicate $r-i\ge 2$ mag for unreddened M giants in the IPHAS survey. 
{    From the VSX catalog \citep{watson16} we got } 24 stars classified as late type variables (L, LB, M, SR, S), only 7 of them with a reported period. All these 24 stars are present in the IPHAS catalog: 6 have color index $1.7<r-i<2.0$ mag and 18 with $r-i>2.0$ mag: the bluest Mira has $r-i=2.22$ mag. 

As our aim is to look for AGB stars with large amplitude variations, which are the reddest ones, we feel confident that a color limit $r-i>1.7$ mag  includes all the {    potential} Mira stars in the field.

The expected variation amplitude of Miras in the $I_\mathrm{C}$ band can be derived from the sample of 154 Mira variables studied by \citet{Maf99} with the same telescope and plates in the field of M16: 95\% of these stars have peak to peak amplitude larger than 2 mag. As the Asiago plates limit is typically around $I_\mathrm{C}=16.5$ mag \citep{Nes16}, we adopted a magnitude cutoff $i=15.0$ mag in the IPHAS catalog to have a high probability to detect our candidate stars also when {    at the faintest magnitude}. Overall 530 IPHAS stars were selected in this way, obviously including the 24 known VSX variables.

Stars brighter than $r=13$, $i=12$ and $H_\alpha$=12.5 {    were found} to be partially saturated in this catalog, so their color indexes may be less reliable. We feel however that our candidate selection is not much affected by this problem. Actually none of our stars has $r<13$, or $H_\alpha<12.5$, while 8 have {    an average} $i<12$.

\section{Photometry}
\label{photom}
The Asiago plates were digitized with an EPSON 1680 Pro scanner at 1600 dpi in transparency mode at the Perugia University \citep{Nes14}, giving a scale of 1.587 arcsec/pixel, while the astrometric solution and automatic photometric reduction were carried out using the PyPlate software\footnote{\url{https://www.plate-archive.org/applause/project/pyplate/}}, a pipeline developed {    by} the APPLAUSE project \citep{Tuv14}.

The pipeline processes digitized plate images in multiple steps. Instru-
mental magnitudes of all detected sources are extracted with the SExtractor
program \citep{Bertin96}  (MAG AUTO magnitudes). Initial astrometric solutions are
derived with the Astrometry.net software \citep{Lang2010}  and refined astrometric calibra-
tion in sub-fields with SCAMP \citep{Bertin2006}. The source list is then cross-matched
with the UCAC4 catalog. For photometric calibration we used the $r'$ and $i'$
(Sloan) passband data from UCAC4. Each photographic plate has a unique
color response that is characterized with the color term C. Magnitudes
in the plate {\it natural}  system, $i_\mathrm{nat}$, are related to the standard magnitudes:
\begin{equation}
    i_\mathrm{nat} = i' + C(r'-i').
    \label{eq:natmag}
\end{equation}
The pipeline calculates reference magnitudes for a series of $C$ values using
Eq.~\ref{eq:natmag} and fits a calibration curve that transforms instrumental mag-
nitudes to the natural system. The accepted $C$ value corresponds to the
smallest scatter of residuals around the calibration curve. The color term $C$
for our plates
was generally close to zero, of the order of $\sim 0.1$. All extracted instrumental
magnitudes were then transformed to the plate natural system. Overall, of
the 530 candidate stars selected only 14 were not actually detected on the
Asiago plates. 

\begin{figure}
\centering
\includegraphics[width=\columnwidth] {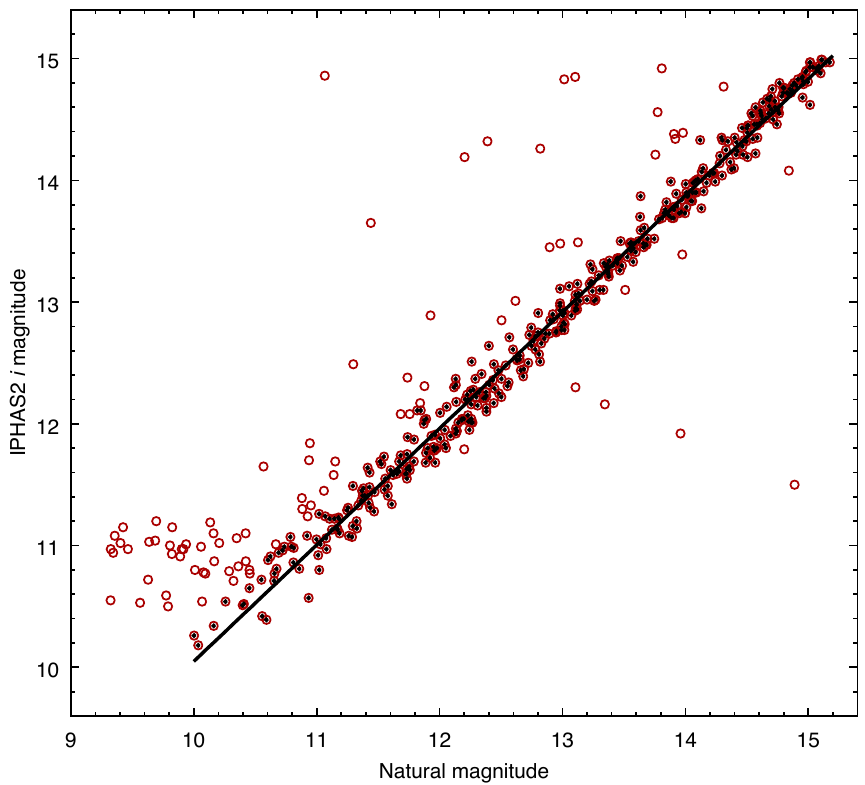}
\caption{Comparison of our "{\it natural} magnitudes" and IPHAS2 $i$ magnitudes for our non-variable red stars {    ( open circles)} in the field of $\gamma$ Cas. Black stars are used to compute the linear correlation described in the text. }
  \label{figure1}
\end{figure}

We report in Fig. \ref{figure1} the comparison between the {\it natural} magnitudes and  the IPHAS $i$ magnitudes for our selected  stars.   For stars fainter than 11 there is a good correlation between {\it natural} and IPHAS magnitudes \\
( iphas=0.96 * {\it natural} + 0.48,  rms=0.12 ).
 The fitting line is also  shown  in the figure.
The photometric accuracy  of our individual magnitudes is typically 0.15 mag, with small variations depending on the plate quality and on the star magnitude.
The accuracy level of the  {\it natural}  magnitudes is good enough to derive meaningful light curves of Miras, while low amplitude variables are more difficult to discover. Also the time sampling, about one plate each month, is mainly suited to search for Mira stars: in this paper therefore we do not attempt to detect small-amplitude variable stars.

\begin{figure}
\centering
 \includegraphics[width=\columnwidth] {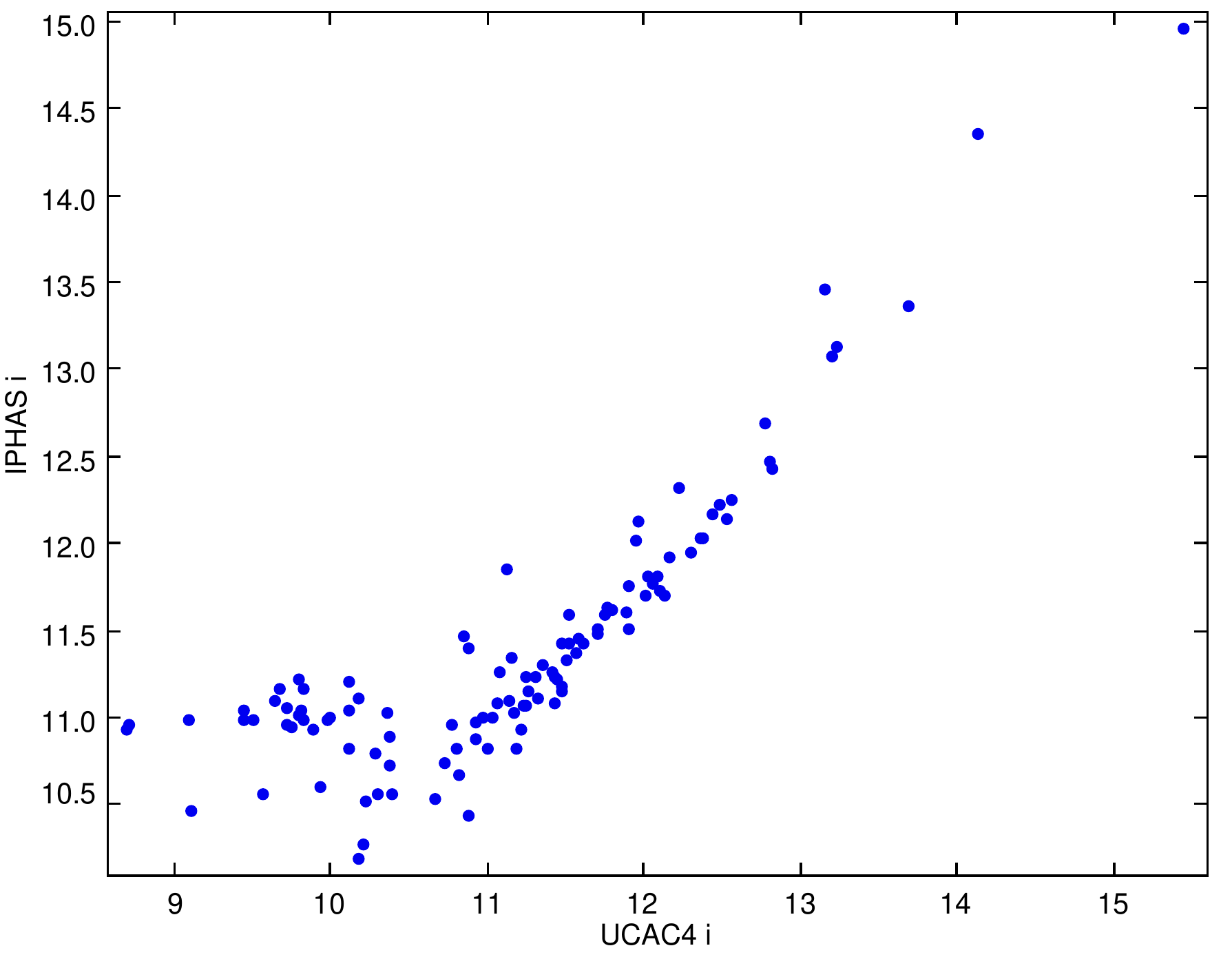}
\caption{Comparison of IPHAS2 $i$ and UCAC4 $i$ magnitudes for our non-variable red stars in the field of $\gamma$ Cas. 
}
\label{figure2}
\end{figure}

A few scattered stars have  magnitudes  quite far  from the  bulk  along the line. This is mostly  due to  coordinate mismatch and  to crowding: {    given the short focal length of the Asiago Schmidt telescope, in some cases one star on a plate actually is a blend of separate IPHAS objects. }
{    On the other hand}, the horizontal strip at the lower left of the plot is due to  saturation  of the IPHAS magnitudes starting at $i$ mag about 12, as also reported in Table 1 of \citet{Bar14}. These problems were further confirmed when we directly compared the IPHAS with the UCAC4 magnitudes of our stars present in both catalogs (see Fig  \ref{figure2}).

\section{Variability check}
\label{variab}
 As a first guess for variability, for each star we examined the rms deviation $\sigma$ from the mean value of the individual magnitudes: most of our stars had  $\sigma$ values between 0.10 mag and 0.30 mag, with a median value 0.18 mag, which is comparable to the typical photometric error of the individual measurements (0.15 mag).  We checked if the observed scatter in measured magnitudes was a function of the mean magnitude: stars fainter than $\sim$11 mag {    had} a well behaved trend, fainter stars {    showed} a larger $\sigma$.

Only 22 stars had $\sigma \ge 0.3$, with 16 larger than 0.5, indicating large photometric variations and therefore possible Mira candidates. One star was excluded because {    it} is very close to the strongly saturated $\gamma$ Cas star, so that its photometry proved to be unreliable, leaving us with 21 candidates.
We checked in the VSX, NSVS  \citep{woz04a,woz04b}  and GCVS \citep{Sam17} databases if any of our variables was already known. Ten stars have a nice match, six stars have no match, five stars have a possible counterpart in the NSVS database but are not present in GCVS or VSX. Often the NSVS database has two or three entries with  very similar coordinates, corresponding to a single real star observed in different runs. The low-amplitude variables listed in VSX were not recovered by our procedure due to {    the applied} cutoff in the scatter of the individual measurements.

\section{Spectroscopic follow-up}
\label{spectra}
Spectroscopic follow-up {    for} the 21 candidate variable stars was {    done} with the Loiano 1.5-m and  the Asiago 1.82-m and 1.22-m telescopes: Loiano spectra were  {    done} with BFOSC using a 2-arcsec slit, at a step of 3.9 \AA; Asiago 1.82-m spectra were  {    done} with AFOSC using a 2.5-arcsec slit and a step of 4.7 \AA; Asiago 1.22-m spectra where taken with a Boller \& Chivens spectrograph at a step of 2.3 \AA.
 The data were reduced by means of standard IRAF procedures~\footnote{IRAF is distributed by the NOAO, which is operated by AURA, under contract with NSF.}.

\begin{figure}
\centering
\includegraphics[width=\columnwidth] {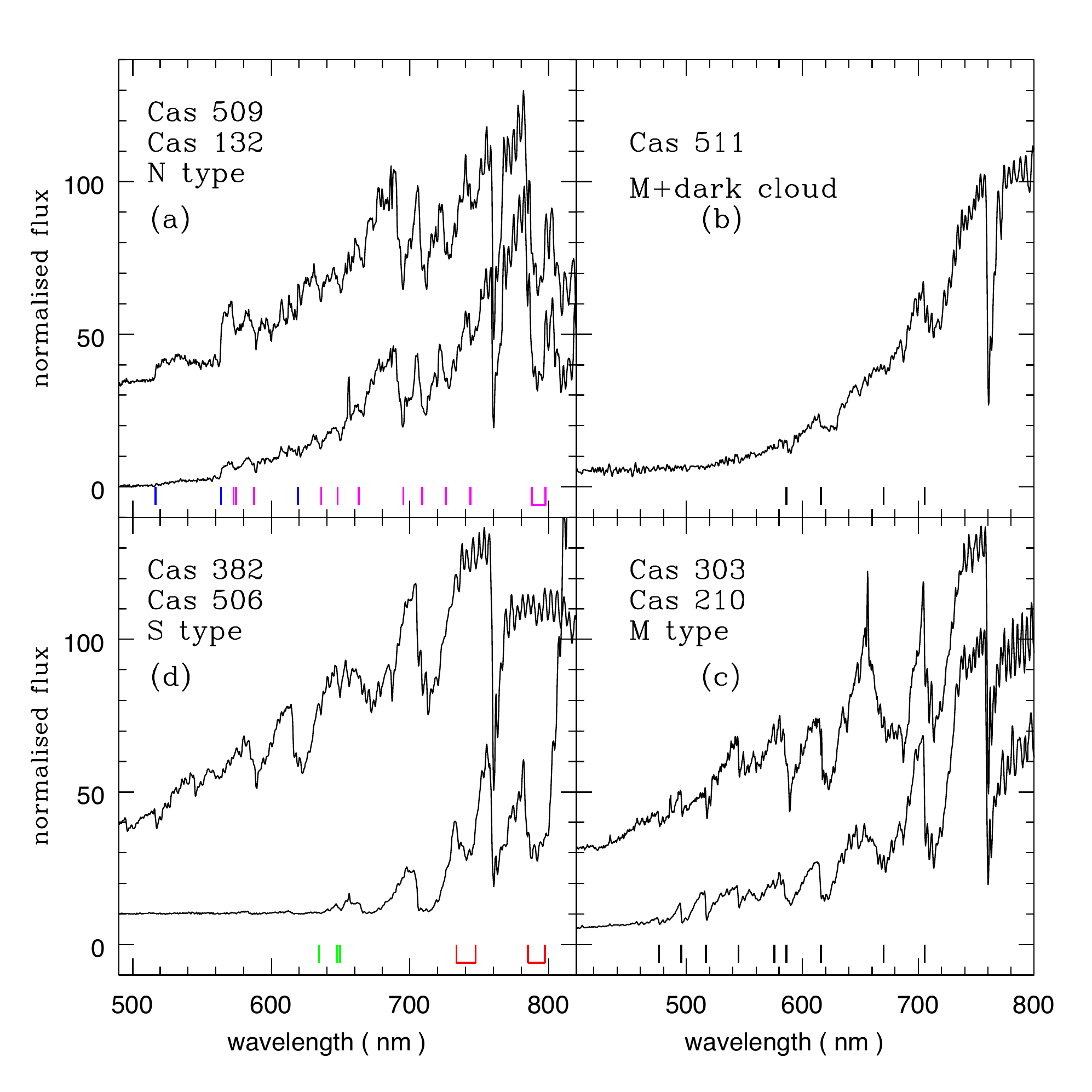}
\caption{ Optical spectra selected among representative stars of  different types.  Vertical dashes indicate the band heads cited in the text: 
~~ (a) blue = C$_2$, magenta= CN; ~(b) and (c) black = TiO; ~(d)  green = ZrO, red = VO.
}
  \label{figurespec}
\end{figure}

{     Information  from literature  regarding the spectral classification of our targets was quite poor; ten stars had no previous classification, the others were classified on the basis of objective prism surveys \citep{Ichi1981, maehara87} or infrared spectra \citep{wright2009}.
 
The aims of the spectral analysis were to discriminate  between different late type stars and the detection of emission lines. {    For} our purposes we  considered the presence of features characteristic of the different spectral types and compared the spectra with those of templates obtained with the same instrumental setup, as in \citet {gaud17a}.
 TiO molecule is present in several spectra
  \footnote { band heads  at 4761, 4954, 5167, 5448, 5759, 5862, 6159, 6700, 7055 and 7600 \AA}. VO bands of the red system are also marginally visible in the later types with the band heads in the range 7334-7472, and 7851-7973 \AA. Six stars show evidence of ZrO molecular absorptions with band heads at 6345, 6474, 6495, 6933 \AA.
For these stars several templates of S type have been taken at Loiano Observatory.

Seven targets are N type AGB carbon stars embedded in a  circumstellar envelope. Their spectra show the C$_2$ Swan system and the red system of the CN molecule, with different depths \footnote{band heads at  5165, 5636 and 6192\AA~ of C$_2$,  and at 5730, 5746, 5878, 6360, 6478, 6631, 6952, 7088, 7259, 7876--7945 \AA of CN.}.
 
Spectral variability is expected for all our stars; the spectral types reported in the tables are for the epoch of our observation. Representative spectra are presented in Fig. \ref{figurespec}.
}

Remarkably, only 3 of our 7 Carbon stars were already present in the Catalog of Galactic Carbon Stars (CGCS, \citealp{Alk01} ). All the M stars, save one (\#303) are of luminosity class III. 

Some of the spectra show hydrogen lines in emission at the epoch of our observation.
Further 15 non-variable stars out of our sample of 530 were also spectroscopically observed and found to be all of late M-types (M5--M8) {    none of them being of C or S-type.}

Direct images in $r_\mathrm{Gunn}$ and $i_\mathrm{Gunn}$ (Loiano) or $r_\mathrm{Sloan}$ and $i_\mathrm{Sloan}$ (Asiago) band were also {    obtained}  for most of the stars to check the magnitude at the epoch of the spectral observation. Comparison stars were taken from UCAC4 catalog, as for the Asiago plates, and photometry was made with IRAF/apphot.

We collect in Table \ref{tab1} positional and spectroscopic data for our 21 stars: column 1 is our internal label, column 2 other identification from VSX or NSVS (if any), column 3 the IPHAS catalog name (i.e. RA+DEC J2000), column 4 the date of observation, column 5 our spectral type,  column 6 the literature spectral type in SIMBAD, if available.

\begin{table*}
\caption{Positions and spectral types of our variable stars.}            
\label{tab1}      
\centering                          
\begin{tabular}{l l l l l l }        
\hline\hline                 
num & other ID &  IPHAS2  & date-obs &  sp.~type & sp.~type  \\
       &                &                 &                &     our     &  SIMBAD \\
\hline\hline                 
091 & NSVS-1638444  & J004531.42+595953.0 & 29 Nov 2016~~ & S7e &M7 \\
098 & NSVS-1668925  & J004551.55+622953.4 & 30 Nov 2016 & N+dust &C \\
132 &  NSVS-1669789 & J004753.07+620245.6 & 29 Nov 2016 & N+dust e &C \\
198 & Dauban-V265    & J005224.44+630342.4 & 29 Nov 2016 & S7e &M8 \\
210 & OT-Cas        & J005302.81+603547.6 & 28 Nov 2016 & M4/5e &M6.5 \\
219 & NSVS-1641101 & J005317.94+623611.4  & 29 Nov 2016 & M7 & M6.5\\
234 &                       & J005351.97+625509.6 & 29 Nov 2016 & N+dust &- \\
297 & V867-Cas      & J005628.30+604709.5 & 30 Nov 2016  & M6e & M7 \\
303 &               & J005644.82+602246.3 & 30 Nov 2016 & M4Ve &- \\
306 & MIS-V1364     & J005701.37+611016.3 & 29 Nov 2016 &  N+dust &- \\
345 & AV-Cas        & J005933.99+604318.4 & 30 Nov 2016 & M5/6e & M8 \\
382 &                & J010215.28+585404.8 & 30 Nov 2016 &  S3/1 & -\\
407 & MIS-V0818       & J010441.97+625255.6 & 30 Nov 2016 & S7e & -\\
431 &                & J010558.90+614348.9 & 30 Nov 2016 & N+dust &- \\
442 & MIS-V1305       & J010642.86+595819.2 & 29 Nov 2016& M7 &- \\
462 & V890-Cas         & J010744.58+590301.9 & 01 Dic 2016 & S4/4 &SX/6e \\
500 & NSVS-1684644 &  J011140.54+601138.0 & 30 Nov 2016 & N+dust &C \\
505 &                &  J011225.70+614146.3 & 25 Jan 2017 & M5/6 &- \\
506 & V418-Cas        &  J011259.79+621046.8 & 29 Nov 2016 & S7/2e &S \\
509 & MIS-V1376      &  J011354.32+603727.2 & 29 Nov 2016 &  N &- \\
511 &               & J011408.65+620418.1 & 14 Feb 2017 & M0?+IS-cloud & -  \\
\hline                                   
\end{tabular}
\end{table*}

\section{Light curves}
\label{lcurves}
A search for periodic variations was made for all our variables using Period04 \citep{Per04}, which is basically a Discrete Fourier Technique (DFT) method. In many cases amplitude variations or long term trends were present which made an automatic period determination difficult. In a few cases the presence of two periods can be identified.

We broadly classified the light curves according to the scheme of the GCVS \citep{Sam17}  in four groups: 
\begin{itemize}
\item { Mira, stars with nearly constant large amplitude and period;}
\item { SemiRegular (SR), stars with large but not constant amplitude and period;}
\item { Trend, stars with a long term trend of the yearly-averaged magnitude;}
\item { Irregular (Irr), stars with irregular amplitude and no clear period.}
\end{itemize}

The difference between Mira and SR in our light curves is not always obvious and is somehow subjective.

\begin{figure*}
\centering
 \includegraphics[width=\columnwidth] {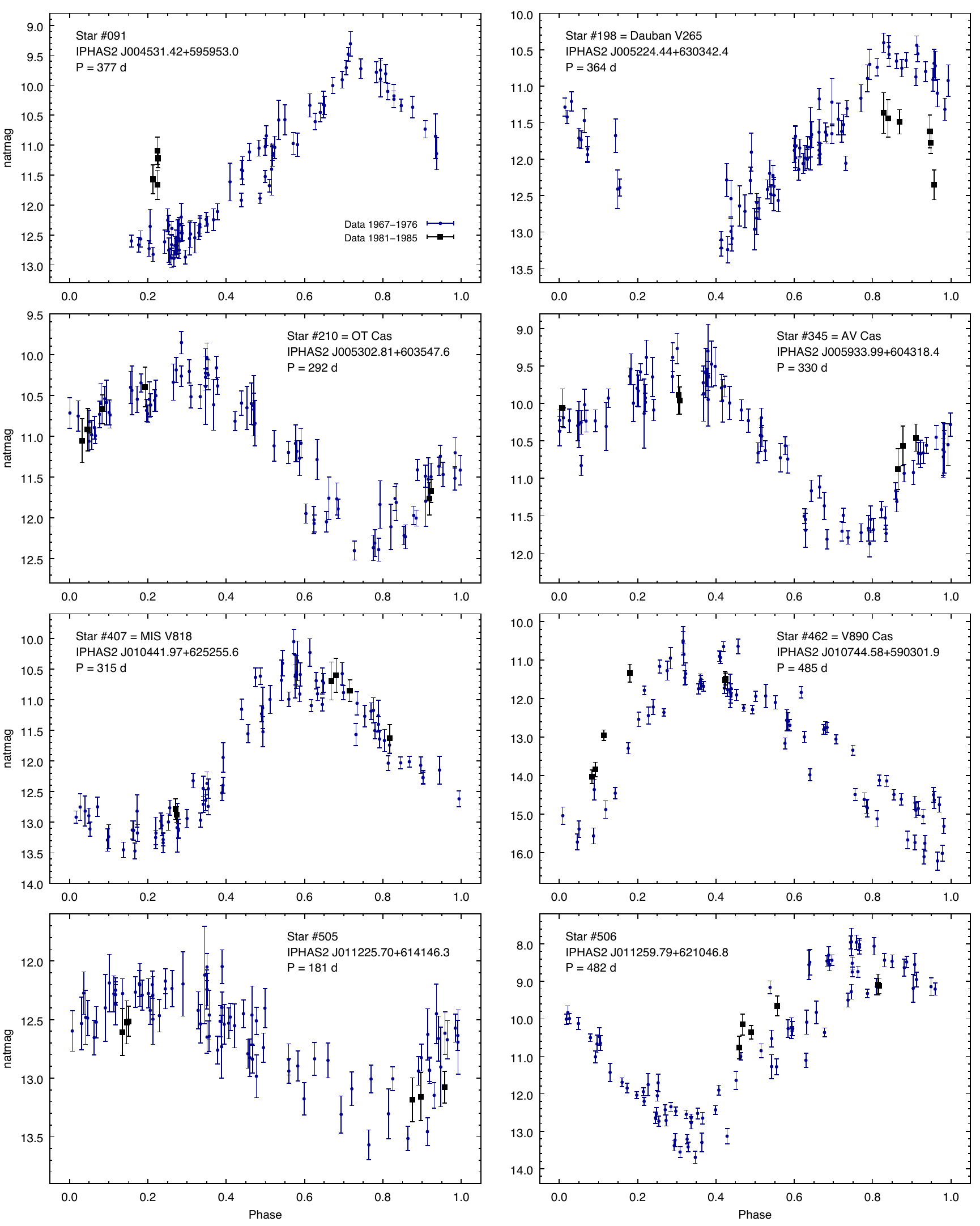}
\caption{Phased Light Curves for Mira stars; black squares are data of 1981--84.}
\label{mira}
\end{figure*}

\begin{figure*}
\centering
 \includegraphics[width=\columnwidth] {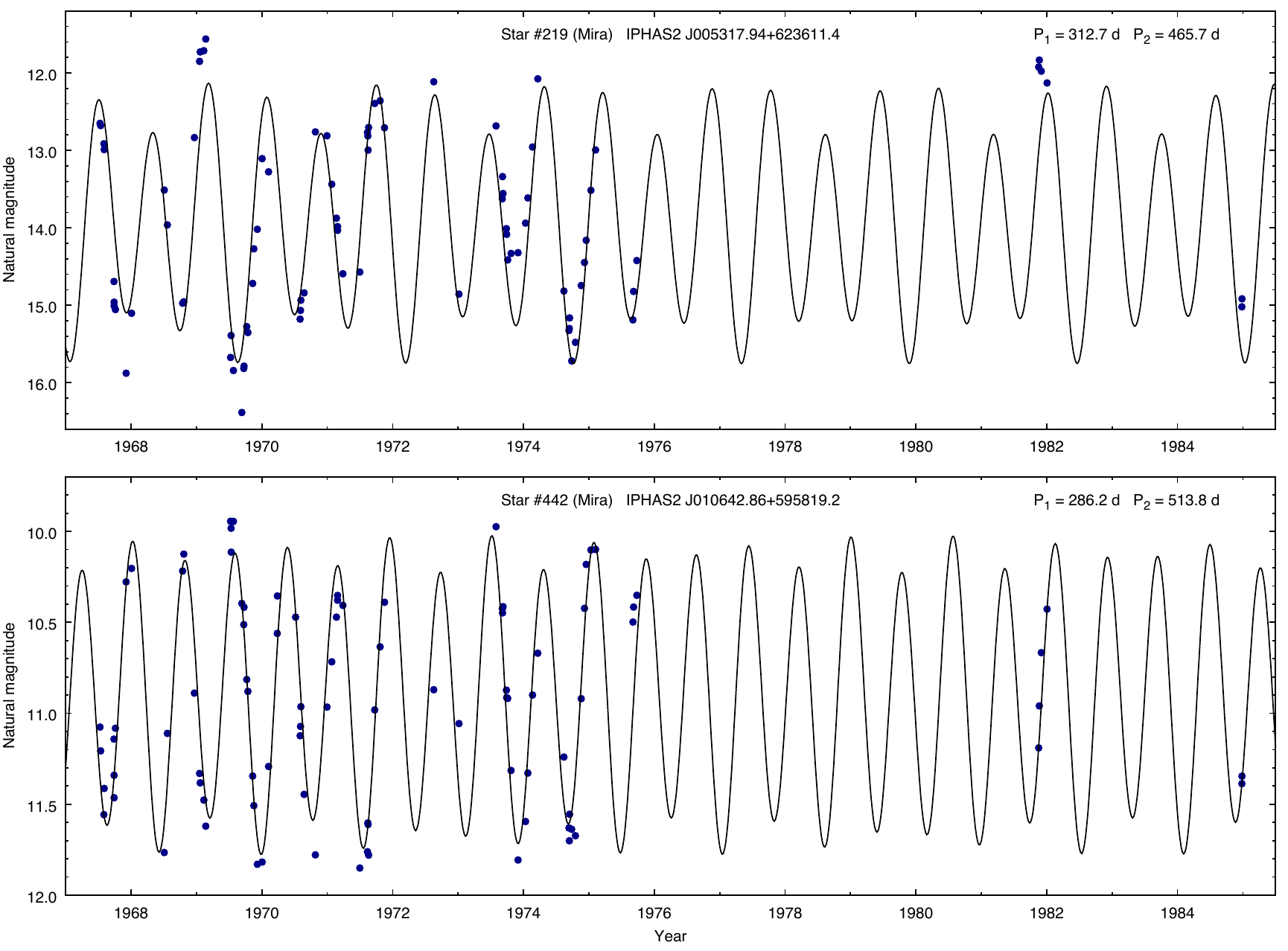}
\caption{Miras with double periods. Upper panel, light curve for star 219 with computed double sinusoidal fit; lower panel, the same for star 442. For both stars the data of years 1981--84 have been used to compute the fit.}
\label{mira2}
\end{figure*}

\begin{figure*}
\centering
\includegraphics[width=\columnwidth] {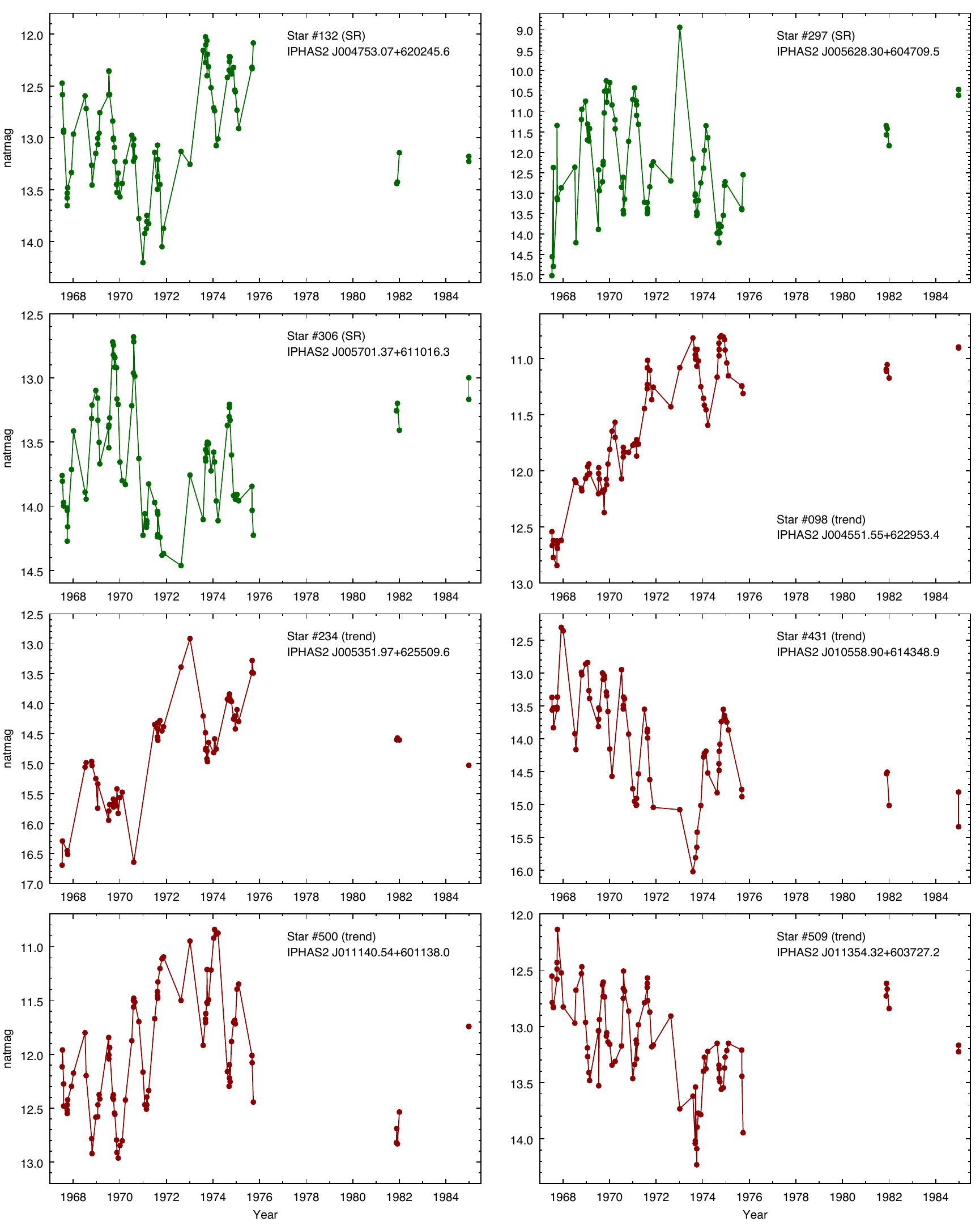}
\caption{Light curves of semi-regular variables (denoted with green points) and variables with long-term trend (red points).}
\label{trend}
\end{figure*}

Overall 10 stars showed a Mira-like light curve, 3  look semiregular, 3 irregulars, 5 stars showed a long-term trend. Phased light curves for the most regular Miras are reported in Fig.~\ref{mira}, light curves for Miras with double period are reported in Fig.~\ref{mira2}, SR stars and stars with clear long-term trend are reported in Fig.~\ref{trend}, Irregular variables are not shown. 

The main photometric data are  compiled in Table \ref{tab2}: column 1 our star label, column 2 our derived period, column 3 the average $i$ magnitude, column 4 and 5 the maximum and minimum recorded magnitudes, column 6 the shape of the light curve, column 7 the spectral type (from Table \ref{tab1}), column 8 the $i$ magnitude at the time of our spectral observation, column 9 the $K$ magnitude from 2MASS \citep{Cut03, skrut06}, column 10 our estimated distance in kpc as discussed in Section 8.

\begin{table*}
\caption{Periods, variation amplitudes and distances of our variable stars. }             
\label{tab2}      
\centering                          
\begin{tabular}{r r r r r r l l r r }        
\hline\hline                 
num & period & $i$-mean & $i$-max & $i$-min & var.~type & sp.~type  & $i$    & $K~~~$      & D \\
        &    days &      mag   &      mag  &    mag   &                  & our          & our &2MASS & kpc \\
\hline\hline                 
091 &  377 & 11.52 & 9.5  & 12.8 & Mira      & S7e&12.1 & 5.40 & 4.5\\
098 &  0     & 11.60 & 10.8 & 12.8 & trend & N+dust&11.2 & 4.82 & - \\
132 &  391 & 12.99 & 12.0 & 14.2 & SR     & N+dust e&12.4 & 5.59 & - \\
198 &  364 & 11.74 & 10.4 & 13.3 & Mira   & S7e & 12.9& 5.53 & 4.4 \\
210 & 292  & 11.07 & 9.8  & 12.4 & Mira    & M4/5e &10.4 & 6.24 & 5.6 \\
219 &  311 & 13.94 & 11.6 & 16.4 & Mira   & M7 & 12.1& 6.72 & 6.6 \\
234 &  -     & 14.69  & 12.9 & 16.7 & trend & N+dust &13.3 &  7.66 & - \\
297 &  380 & 12.34 & 9.0  & 15.0 & SR    & M6e & 8.9 & 4.87 & - \\
303 &  -      & 14.33 & 13.4 & 15.0 &  irr    & M4Ve & - &11.56 & - \\
306 &  294 & 13.62 & 12.7 & 14.4 & SR   & N+dust & 14.1 &7.58 & - \\
345 &  330 & 10.43 & 9.3   & 11.8 & Mira   & M5/6e & 9.8 & 5.67 & 4.9 \\
382 &  -      & 11.84 & 10.8 & 12.5 & irr       & S3/1 & 12.1 & 8.24 & - \\
407 & 315  & 11.92 & 10.1 & 13.4 & Mira  & S7e & 12.5 & 5.41 & 3.8 \\
431 &  -      & 13.97 & 12.4 & 16.0 & trend & N+dust & - & 6.93 & - \\
442 &  287 & 10.94 & 9.9   & 11.9 & Mira  & M7 & 10.1& 5.47 & 3.8 \\
462 &  485 & 13.15 & 10.5 & 16.1 & Mira & S4/4 &12.3& 5.46 & 5.8 \\
500 &  400 & 11.99 & 10.9 & 13.9 & trend & N+dust&12.1 & 5.29 & - \\
505 &  181 & 12.61 & 12.1 & 13.5 & Mira  & M5/6 &12.0& 7.99 & 8.1 \\
506 &  482 & 10.63 & 8.0   & 13.6 & Mira   & S7/2e&11.7 & 3.66 & 2.5 \\
509 &  343 & 13.10 & 12.2 & 14.2 & trend & N & 13.0& 8.13 & - \\
511 &  -      & 15.01 & 14.5 & 15.5 & irr     & M0?+IS &15.3& 8.01 & - \\
\hline                                   
\end{tabular}
\end{table*}

Most of the Mira and Semiregular variables have periods between 290 and 400 days: two S-type stars have periods of about 480 days (V418 and V490 Cas)  while one star (\#505)  has a short period of only 181 days. All stars  that show long-term trends in their light curves are carbon stars.

Below we briefly give some details on the individual stars. The time coverage of the light curves is rather uniform for the first 7 years, although undersampled in 1972/73, and is well suited to derive an average period, if present; the few data {    points} obtained in the 1980's are most useful to check the constancy in period (and/or) phase of the light curve.

\medskip
{\bf Mira stars:}

Star091 (Fig.~\ref{mira}): the star has a period of 377\,d with $\sim$2 mag peak to peak amplitude. The points of the years 1982--84 are out of phase, suggesting a phase shift or a small period change in the years after 1975. The NSVS catalog has three sources very near to our position, \#1638444, \#1587497, \#1665936. 

Star198 (Dauban V265, Fig.~\ref{mira}): a period of 365\,d fits quite well the light curve between 1967 and 1975: the later data of 1982--84 are out of phase by 0.5, indicating a probable phase shift or period change. No period is given in VSX. Possible counterparts in NSVS are \#1639949 and \#1673661.

Star210 (OT Cas, Fig.~\ref{mira}): the period of 292\,d fits quite well the whole dataset. The star is classified as Mira with $P=288$\,d in VSX, in a very good agreement with our result, and with the light curves of the NSVS sources \#1643267 and \#11671838. 

Star219 (NSVS \#1641101, Fig.~\ref{mira2}): a period of 312\,d is apparent, modulated by a longer period of 465\,d in 3/2 ratio, with amplitudes 1.45 and 0.36 mag respectively. NSVS  \#1673811 is the same star.

Star345 (AV Cas, Fig.~\ref{mira}): the period of 330\,d is overall good on the whole time interval. Classified as Mira with $P=322$\,d in VSX, in good agreement with our findings. Possible NSVS counterparts are \#1647501 and \#1676514.

Star407 (MIS V0818, Fig.~\ref{mira}): a good fit is given by a period of 315\,d in the whole time interval 1967--84. No period in VSX. NSVS counterpart is \#1647990.

Star442 (MIS V1305, Fig.~\ref{mira2}): the light curve of this Mira star is well fitted by  a double sinusoid, with periods 287\,d (amplitude 0.76 mag) and 515\,d (amplitude 0.11 mag). No period reported in VSX. NSVS counterparts are \#1729761 and \#1681126.

Star462 (V890 Cas, Fig.~\ref{mira}): this star was studied as a test case for the Asiago plates by \citet{Nes16}, who made a detailed discussion of its light curve. The phased light curve shows a good fit with an average period of 485\,d but appreciable amplitude variations are present in different cycles. \citet{Nes16} explored the trick to double the formal period to see if the phased light curve looks better but without convincing results. No counterpart is reported in NSVS. 

Star505 (Fig.~\ref{mira}): the period of 181\,d seems quite stable but the amplitude is not constant. It is by far the shortest period variable in our sample. \citet{wmf00} divide the Mira with period below 225\,d in two groups, ``Short Period$-$Blue'' and ``Short Period$-$Red''  depending on their infrared colors and average spectral types. The spectroscopic (M6) and photometric (NIR colors) characteristics lead us to place this star in the  group of the ``Short Period$-$Red'' Miras. No counterparts in NSVS.

Star506 (V418 Cas, Fig.~\ref{mira}): a period of 482\,d gives a good fit to the data, but the amplitude seems not constant. We tried to double the adopted period, as in the case of V890 Cas, but the phase plot gives an equally good fit. A period of 480\,d is reported in VSX in full agreement with our findings. NSVS counterpart is \#1686590.

{\bf Semiregular stars:}

Star132 (Fig.~\ref{trend}): a period of 391\,d gives a fair overall match to the data, but marked amplitude variations are present. The sum of two periods, 388\,d and  5.7 years, with similar amplitudes gives a better fit, but not fully satisfactory. The star is present as \#117 in CGCS but with 13-arcsec offset, and without a spectral classification. A possible match candidate is NSVS \#1669789. Another star is at 4 arcsec distance but is 2 mag fainter in $i$, so the Asiago light curve is not much affected.

Star297 (V867 Cas, Fig.~\ref{trend}): a period of 380\,d reproduces a few variability cycles, but both period and amplitude changes are apparent in the light curve. Faint ($I=16$) minima were recorded in 1967 and 1974 and a very bright maximum ($I=9$) in 1973. A longer period of 410\,d, with large uncertainty, was reported by \citet{naka00} but is not reported in VSX. No counterparts are present in NSVS.

Star306 (MIS V1364, Fig.~\ref{trend}): This star is an optical double at 2" distance and therefore it is not resolved in the Asiago plates. Both components are present in the IPHAS catalog, the redder one being our N-type variable. A period of 294\,d describes the overall behaviour but the amplitude is clearly not constant in different cycles, ranging from 1 to 2 mag, peak to peak. The presence of the stable star ($i=14.44$) within the PSF of the Asiago plates limits the actual variability range detected. No period is given in VSX. No counterparts in NSVS. Not listed in CGCS.

{\bf Trend stars:}

Star098 (Fig.~\ref{trend}): the star is listed in the CGCS (\#110), but without a spectral classification. Counterparts in NSVS are \#1669789 and \#1636119.

Star234 (Fig.~\ref{trend}): the star has small amplitude variations and a long term trend of its average level. This star has  no counterparts in NSVS and is not listed in CGCS.

Star431 (Fig.~\ref{trend}): a time scale of about 330\,d is apparent, superimposed on a long term trend of the mean flux. An inversion of this trend after 1974 is also present. No counterparts in NSVS. Not listed in CGCS.

Star500 (Fig.~\ref{trend}): a long term oscillation is superimposed on the main period of $\sim$400\,d. This star is present in the KISO catalog (KISO C1-137)  of carbon stars and is listed in the CGCS (\#183), but again without a spectral classification. NSVS counterparts are \#1684644 and \#1733537.

Star509 (MIS V1376, Fig.~\ref{trend}): the star showed a long term monotonic decreasing trend form 1967 to 1975, with wide oscillations of about 1 magnitude, but without a clear period. In the years 1982--85 the magnitude of the 
 star was again at the level of 1967, indicating an inversion of the long term trend. Classified SR in VSX. A possible counterpart in NSVS is \#1686429 but with only 180 days coverage and scattered points with small amplitude variability. Not listed in CGCS.

{\bf Irregular stars:}

Star303: the light curve shows a 1 mag peak to peak amplitude.
 The spectrum is the only one of Luminosity Class V  in our sample. It shows (see the bottom right plot of Fig \ref{figurespec}) very strong Balmer lines in emission from H$\alpha$ ~to H$\gamma$; {    to understand the origin of these features and of the luminosity variability a denser photometric and spectroscopic monitoring with higher resolution would be necessary}. Its proper motion is 20 mas/yr in the URAT1 catalog, or 45 mas/yr  in UCAC5, parallel to the galactic plane, of the expected order of magnitude for a relatively nearby dwarf. {    The Gaia DR2 catalog, published while this paper was under revision, gives a distance of 185 pc (parallax 5.41 mas) and a proper motion of 26 mas, confirming its dwarf nature.}

Star382: this is the only irregular variable among our S-type stars and it is by far the hottest of them.

Star511: this star has a few much scattered points in the plates of 1967 and one in 1970; these points produce the observed large rms deviation from the mean: excluding these points, the star shows small variations, comparable with the photometric uncertainty. It is behind the dark cloud LDN 1312, which  affects the color indices and hampers a reliable classification of spectral type and luminosity class. The few, strongly veiled, absorptions  and 
  the absence of molecular bands in the reddest part of the spectrum  suggest to assign to this star a  spectral type  between late K and early M.

\section{Color-color plots}
\label{colcol}

\begin{figure}
\centering
  \includegraphics[width=\columnwidth] {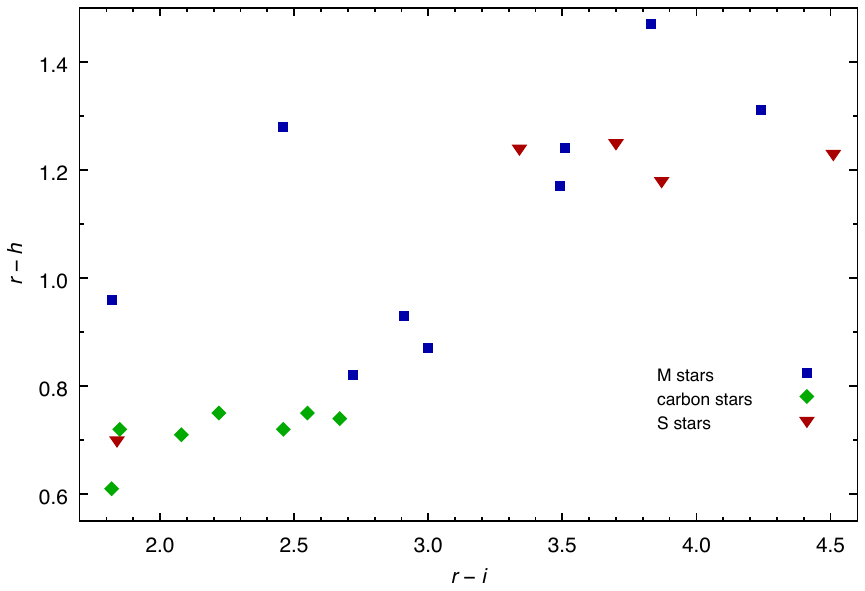}
\caption{The $r-h$ {\it vs} $r-i$ color plot for our variables. Spectral types are coded with different symbols: squares = M stars; diamonds = carbon stars; triangles = S stars. 
{    Most of the stars are distributed along a diagonal path from lower left to upper right.
The S-type star at the lower left corner is the Irregular variable \#382; the two M-type stars in the upper left part of the plot are \#210 and \#303}.}
\label{figure3}
\end{figure}

The plot of the original IPHAS2 $r-h$ {\it vs} $r-i$ colors for our variables is shown in Fig \ref{figure3}: H$\alpha$ emission line stars are expected to have a large $r-h$ color and cooler stars a larger $r-i$.
Our M-type giant stars, ranging from M5 to M7, are distributed all along the diagonal of this plot. 
Our rather late (S7) S-type stars are in the upper-right area, while  the hot S3 star \#382 lies in the lower left corner, in the zone occupied by the  dusty, strongly variable C-type stars.
 Two stars are worth of a short mention:

1) star \#303: {    the IPHAS2 colors ($r-i=1.85, \, r-h=0.97$) put this star in the locus of the unreddened ones (see figures 15-17 of \citealp{Bar14}); together with its spectral classification as a Main Sequence star, this implies a short distance and a low interstellar absorption}. 

2) star \#210: its colors  ($r-i = 2.5, \, r-h = 1.3$) {    put} the star in the region of strong $H\alpha$ emitters. Our spectrum shows a faint $H\alpha$ emission, but H$\beta$, \,H$\gamma$  ~and  H$\delta$ \, are present. The IPHAS $r$ and H$\alpha$ ~magnitudes are not saturated (see Table 1 in  \citealp{Bar14}  for threshold values), while $i$ may be partially affected. Our measure of $r-i$, taken when the star was in a {    bright} state (see Table 2), is 2.1 mag, significantly bluer than the IPHAS2 value.
 
{    Overall, the spectral types of our stars are rather scattered in this plot, so it is of little use for a reliable spectral classification. Only obscured carbon stars are segregated in a well defined area: therefore the IPHAS colors, in connection with infrared (JHK) plots, are very useful to select this type of stars with sufficient confidence level.}

\begin{figure}
\centering
  \includegraphics[width=\columnwidth] {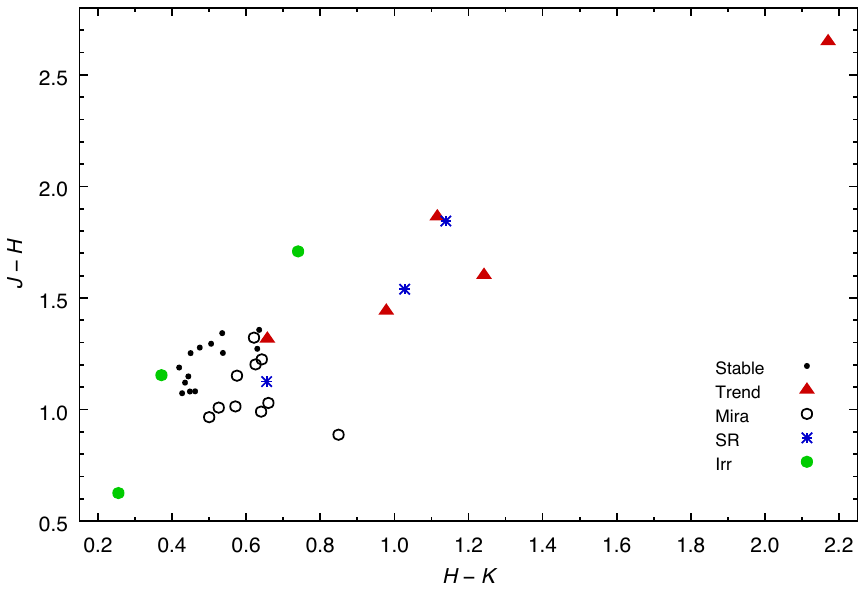}
\caption{The $J-H$,$H-K$ diagram of our variables.  The Carbon  star \#234 is located at the upper-right corner; the main sequence star \#303 is at the lower left corner. {    
Variability type is coded as follows: open circles = Mira; blue  stars = semiregular variables; filled  red triangles= stars with a long term trend; filled green circles = irregular variables; dots = non-variable (stable) stars.} }
\label{figure4}
\end{figure}

We cross-correlated the positions of our detected IPHAS stars with the 2MASS and WISE \citep{Cut13} catalogs, in order to build color-color plots and look for useful correlations: nearly all of our 530 stars have a counterpart in these catalogs. 
We plot in  Fig. \ref{figure4} the $J-H$,$H-K$ diagram of our variables, together with our non-variable stars with determined spectral type: the separation of Mira, SR and Irr is not well defined {    in this plot}. Most carbon stars are easily discriminated by their very red colors. M-type Mira variables appear on average below the non-variable stars, a result already found e.g. in \citet{Men85},\citet {Whi94} and recently in \citet{gaud2} for high galactic latitude stars.
The irregular variable in the lower left corner is the Main Sequence star \#303: its position is consistent with the expectation from the \citet{BeB88} color-color diagram.

Our stars are quite bright (w1$<$8 mag) in the WISE catalog and therefore possibly saturated (see \citealp{Nik14}):  we explored anyway several color-color plots, but they did not add significant informations for our stars.

We found no clear trend between any color index and the period length for any kind of variables.

\section{Distance estimates}
\label{distance}
\begin{figure*}
\centering
  \includegraphics[width=\columnwidth] {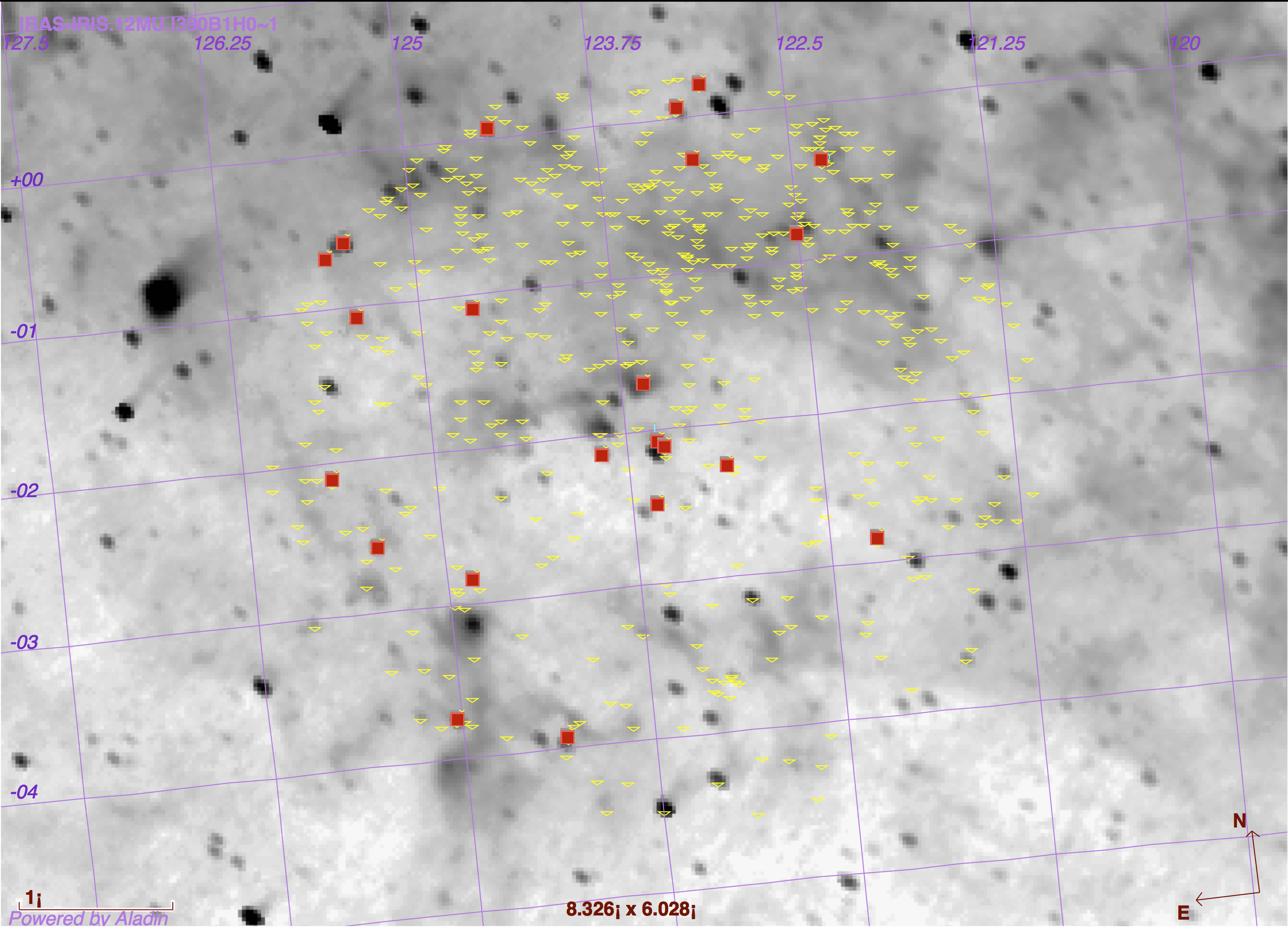}
  \caption{The emission map (negative) at 12 microns  of  IRAS centered on $\gamma$ Cas. The stars with $r-i>1.7$ mag are shown as (yellow) triangles, our variables are (red) squares.}
\label{figure5}
\end{figure*}

The sky area around $\gamma$ Cas is near the galactic plane, so that strong and inhomogeneous absorptions in the optical and NIR bands are expected. An indication of the existence of such absorptions is given by the IRAS 12$\mu$ emission map, shown in Fig. \ref{figure5} overplotted to our 530 stars. A prevalence of stars in the upper part of the figure, richer of cold emission features, is evident, but this is not true for our variable stars which are distributed more evenly in Dec.

We have tried to estimate the distances for the Mira stars; For this purpose we computed their absolute $K$ magnitudes using the relation given by \citet{white12}, {    based on a sample of Miras with parallaxes measured by HIPPARCOS}:  

\begin{equation}
    M_K= -3.69 (\log P - 2.38) - 7.30.
\end{equation} 
Then the distance $D$ was computed from
\begin{equation}
    \log D = \frac{(K - M_K + 5 - A_K)}{5},
\end{equation}
$A_K$ being the absorption in front of the star.

Given the variability of a Mira star, its 2MASS $K$ magnitude is somewhat different from its average value, which should {    actually} be used to derive the distance. The peak-to-peak amplitude in this band is about 0.6 mag \citep{wmf00} with only a few stars reaching 1 full magnitude amplitude, {    so a reasonable estimate of the uncertainty on M$_K$ is 0.3 mag.} 
 
An estimate of the absorption along the line of sight can be attempted in different ways. A recent evaluation of a 3D absorption map in our region was made by \citet{Sal14} on the basis of the IPHAS survey with a nominal resolution of 10 arcmin$^{2}$.
We used this to estimate the absorptions of all our 21 variables assuming a 
distance of 2.3 kpc (the Perseus Arm), finding $A_V$ values ranging from 1.0 to 3.1 (mean value 1.86): this corresponds to a range in $A_K$ from 0.1 to 0.3 \citep{Sch98}.
 
An independent way to estimate the absorption is from the observed $J-H$,$H-K$ colors of our stars, assuming as intrinsic values the average of the Miras of high galactic latitude in the GCVS. Mira stars occupy a well defined region in the $J-H$,$H-K$ diagram (see e.g. \citealp{Whi94}) centered at  $J-H=0.91$, $H-K=0.53$: the mean colors of our Miras are $J-H=1.07$, $H-K=0.62$, giving a color excess $E(B-V)=0.62$ and a $K$-band absorption of 0.23 mag, in good agreement with the \citet{Sal14} estimate.

A further estimate of the absorption for each Mira star can be {    made} using the relation between the period and the intrinsic $(J-K)_0$ color according to \citet{wmf00}:
\begin{equation}
    (J-K)_0=-0.39+0.71 \log P.
\end{equation}

In any case the expected absorption in $K$ band is at most a few tenth of magnitude, comparable to, or smaller than, the uncertainty of the actual mean value of the $K$ apparent magnitude for each star.

To search for evidence of dust near our stars, we made a careful inspection of the WISE $w3$ (11.6$\mu$) images for all the variables reported in Table \ref{tab2}: only 6 stars showed nearby nebulae and are reported in Table \ref{WISE}. In four cases the stars look to be outside the apparently nearby nebulae, while only one of them (\#505) looks embedded in substantial emission. Its position on the sky in Fig. \ref{figure5} is at galactic coordinates l=125.42 b=-1.07.

\begin{table}
\caption{Variable red stars with surrounding/nearby emission at 11.8$\mu$}             
\label{WISE}      
\centering                          
\begin{tabular}{l l l l}        
\hline\hline                 
num & sp.~type & var & nebula \\
\hline\hline                 
132 &N+dust & SR   & nebula at NE\\
219 & M7       & Mira & faint diffuse emission\\
290 & M5       & Irr     & diffuse light from $\gamma$ Cas \\
306 & N+dust & SR & strong emission to E and S\\
505 & M5/6    & Mira & substantial bkg emission \\
509 & N         &Trend & substantial emission around\\
\hline                                   
\end{tabular}
\end{table}

 {    From equation 3, a variation of 0.4 mag gives a variation of about 20\% in the computed distance: the combined uncertainty on actual mean K value and interstellar absorption on average should not be larger than 0.4 mag, so we guess that our estimated distances have an uncertainty of $\sim$20\%.}
 
 Adopting the absorption estimate from the $J-K$ excess (equation 4), we derived the distance for each Mira which is reported in the last column of Table \ref{tab2}: these distances range from 2.5 to 8.1 kpc, with a median value of 4.9 kpc. We found no clear correlation of the star distances with their position in the sky.

{    The Gaia DR2 astrometric catalog (\citet{Gaia};  \citet{Lin18}) was published after the paper was submitted for publication. All our Miras are present in the Gaia database, with  errors on the parallax ranging of 0.08 to 0.27 mas and median value 0.13 mas.
This means an accuracy of 13\% if the distance is 1 kpc but only $\sim$50\% at 4.5 kpc, which is the median distance of our stars estimated  from the Period-Luminosity relation. 
In particular, for 4 stars (\#091, \#219, (\#407, \#462) the Gaia parallax has an error comparable to the formal value,  for 2 stars (\#210, \#505) has an uncertainty about 40\%, and for 4 stars (\#198, \#345, \#442, \#506)  between 20\% and 30\%. 
Furthermore, many parameters describing the overall quality of the Gaia astrometric solution of our stars (most notably astrometric\_gof\_al, astrometric\_excess\_noise, astrometric\_weight\_al) are quite bad, so the actual uncertainties on the parallaxes are surely significantly larger than the formal values.

Under these conditions, also for our 4 best cases the statistical meaning of the distance derived as inverse of the parallax is very ill-defined (see \citet{Bailer18}, and \citet{luri18}), so that a comparison of the photometric and trigonometric parallax for a given star is not very meaningful.
}

\section{Conclusions}
\label{sect:conclude}

We found 21 large amplitude variables in our sky region: 10 Miras, 3 Semiregular, 3 irregulars and 5 with long term trends. The two stars with the longest periods, about 480 days, have an S-type spectrum. Only one short period (180 days) Mira was found. 

Just 13 of these stars were already listed in the VSX or GCVS catalogs of variable stars, but only 4 of them had a published period: our search has therefore substantially improved the number of large amplitude red variables known in this sky area, and provided robust period determinations for all of them, likely completing the sample at our magnitude level.

For  comparison, we refer to the result of a similar search made on plates taken with the same telescope and emulsion \citep{Gas91} in a nearby 5x5 degrees field centered on IC 1805 (l=134.5, b=+1.0). These authors found 28 red variables, 12 of them being Miras: given the similarity of the galactic region explored, the number of Miras found in the two fields is  consistent, assuming a Poisson statistics uncertainty.

We remark that all (and only) the N-type stars with strong infrared excess showed long term trends in their light curves.
Our time base is too short to establish if any periodicity exists in these long term trends. 
A similar result was found, for instance, by \citet{Whi03} (and references therein) in the LMC.
If these trends are erratic, the most reasonable interpretation is  a mass loss variation and the presence of circumstellar envelopes of variable density, producing a variable dimming and brightening without a definite periodicity.
A dynamical model approach to this problem may be found in \citet{Hof16} and references therein; recent applications to carbon stars atmospheres may be found in \citet{Rau17} and references therein.

{    Finally, we note that, rather surprisingly, most of our Mira variables have a photometric distance beyond the Perseus Arm. 
Overall, we feel that a measure of the actual distances of our Mira stars must wait for future improved releases of the Gaia catalog after the end of the scheduled 5 years of observations, and for possible improvements of the Period-Luminosity relation based on more nearby Miras with high-quality Gaia parallaxes. 
}

\begin{acknowledgements}
{\bf Acknowledgements.}

{    We thank Beatrice Bucciarelli for useful hints to understand the numerous quality parameters of the Gaia DR2 catalog.}

This research has made use of the SIMBAD database, operated at CDS, Strasbourg, France;  
of the Two Micron All-Sky Survey database, which is a joint project of the University of Massachusetts and
the Infrared Processing and Analysis Center/California Institute of Technology; 
of NASA's Astrophysics Data System Bibliographic Services.
This paper makes use of data obtained as part of the INT Photometric H-alpha Survey of the Northern Galactic Plane (IPHAS)
carried out at the Isaac Newton Telescope. 
All IPHAS data are processed by the Cambridge Astronomical Survey Unit.
 The band-merged DR2 catalogue was assembled at the Centre for Astrophysics Research,
University of Hertfordshire, supported by STFC grant ST/J001333/1.
The Asiago plates were processed with the PyPlate software at AIP, 
supported by the Deutsche Forschungsgemeinschaft (DFG) grant EN 926/3-1.
TT acknowledges support by the Centre of Excellence $Dark\, side\, of\,  the\, Universe\, $ (TK133) financed by the European Union through the European Regional Development Fund and  by the institutional research funding IUT26-2 IUT40-2 of the Estonian Ministry of Education and Research.

Finally we aknowledge the use of the Gaia-DR2 catalog.
\end{acknowledgements}




\begin{thebibliography}{}
\bibitem[Alksnis et al.(2001)]{Alk01} Alksnis, A., Balklavs, A., Dzervitis, U., et al. 2001, Baltic Astronomy, 10, 1
\bibitem[Bailer-Jones et al. (2018)] {Bailer18}Bailer-Jones,C.A.L., Rybizki, J.,  Fouesneau, M.,  et al. 2018 arXiv:1804.10121, ApJ in press.
\bibitem[Barentsen et al.(2014)]{Bar14} Barentsen, G., Farnhil, H. J., Drew, J. E., et al. 2014, MNRAS, 444, 3230
\bibitem[Bertin \& Arnouts(1996)]{Bertin96} Bertin, E., \& Arnouts, S. 1996, A\&AS, 117, 393
\bibitem[Bertin(2006)]{Bertin2006} Bertin, E. 2006, in Astronomical Data Analysis Software and Systems XV, ed. C. Gabriel, C. Arviset, D. Ponz, \& S. Enrique, Astronomical Society of the Pacific Conference Series, 351, 112
\bibitem[Bessell \& Brett(1988)]{BeB88}Bessell, M.S. \& Brett, J.M. 1988, PASP 100, 1134
\bibitem[Cutri et al.(2003)]{Cut03} Cutri, R. M., Skrutskie, M. F., van Dyk, S., et al. 2003, The 2MASS All-Sky Point Source Catalog, University of Massachusetts and Infrared Processing and Analysis Center (IPAC/California Institute of Technology), VizieR On-line Data Catalog: II/246 
\bibitem[Cutri et al.(2013)]{Cut13} Cutri, R. M., et al. 2013, WISE All-Sky Data Release, IPAC/Caltech, VizieR On-line Data Catalog: II/328
\bibitem[Gaia coll.(2018)]{Gaia} Gaia collaboration, CDS catalog I/345/gaia2
\bibitem[Gasperoni et al.(1991)]{Gas91} Gasperoni, V., Maffei, P., Tosti, G. 1991, IBVS, 3573
\bibitem[Gaudenzi et al.(2017a)] {gaud17a} Gaudenzi, S., Nesci, R., Rossi, C., Sclavi, S.,  Gigoyan, K. S.,  Mickaelian, A. M. 2017a, RMxAA, 53, 449
\bibitem[Gaudenzi et al.(2017b)] {gaud2} Gaudenzi,  S., Nesci, R., Rossi, C., Sclavi, S.,  Gigoyan, K. S.,  Mickaelian, A. M. 2017b, RMxAA, 53, 507
\bibitem[Hofner et al.(2016)]{Hof16} Hofner, S., Bladh, S., Aringer, B., Hauja, R. 2016, A\&A 594, 108
\bibitem[Ichikawa (1981)]{Ichi1981} Ichikawa, T., 1981 PASJ, 33, 107
\bibitem[Lang et al.(2010)]{Lang2010} Lang, D., Hogg, D. W., Mierle, K., Blanton, M., \& Roweis, S. 2010, AJ, 139, 1782.
\bibitem[Lenz \& Breger(2004)]{Per04} Lenz, P., \& Breger, M. 2004, IAUS, 224, 786.
\bibitem[Lindegren et al. (2018)]{Lin18} L. Lindegren, J. Hernandez, A. Bombrun, A., et al. 2018 A\&A in press.
\bibitem[Luri et al. (2018)]  {luri18}  Luri, X., Brown, A., Sarro, L., et al. 2018, arXiv:1804.09376, A\&A in press
\bibitem[Maffei (1977)]{Maf77}Maffei, P., 1977, IBVS, 1302
\bibitem[Maffei \& Tosti(1999)]{Maf99} Maffei, P., \& Tosti, G. 1999, Pub. Univ. Perugia, VizieR On-line Data Catalog: II/320
\bibitem[Maehara \& Soyano(1987)]{maehara87} Maehara, H., Soyano, T., 1987 ,  Ann. Tokyo Astron. Obs., 21, 293-310.
\bibitem[Menzies \& Whitelock(1985)]{Men85}Menzies, J.W., \& Whitelock, P.A. 1985, MNRAS 212, 783.
\bibitem[Nakashima et al.(2000)] {naka00} Nakashima, J., Jiang, B. W., Deguchi, S., Sadakane, K., \& Nakada, Y. 2000, PASJ, 52, 275 
\bibitem[Nesci et al.(2014)]{Nes14} Nesci, R., Bagaglia, M., Nucciarelli, G. 2014, in Astroplate 2014, ed. L. Mi{\v s}kov\'a \& S. V\'itek (Institute of Chemical Technology, Prague), 2014, 75
\bibitem[Nesci(2016)]{Nes16} Nesci, R. 2016, IBVS, 6170, 1
\bibitem[Nikutta et al.(2014)]{Nik14}Nikutta, R., Hunt-Walker, N., Nenkova, M., et al. MNRAS 442, 3361. 
\bibitem[Rau et al.(2017)]{Rau17} Rau, G., Hron, J., Paladini, C., et al. 2017, A\&A, 600, 92
\bibitem[Reid et al.(2014)]{Rei14} Reid, M. J., Menten, K. M., Brunthaler, A., et al. 2014, ApJ, 783, 130
\bibitem[Sale et al.(2009)]{Sal09} Sale, S. E., Drew, J. E., Unruh, Y. C., et al. 2009, MNRAS, 392, 497
\bibitem [Sale et al.(2014)]{Sal14} Sale, S. E., Drew, J. E., Barentsen, G., et al., 2014, MNRAS, 443, 2907
\bibitem[Samus et al.(2017)]{Sam17} Samus, N. N., Kazarovets, E. V., Durlevich, O. V., Kireeva, N. N., \& Pastukhova, E. N. 2017, General Catalogue of Variable Stars: Version GCVS 5.1, Astronomy Reports, vol. 61, No. 1, pp. 80-88, VizieR On-line Data Catalog: B/gcvs
\bibitem[Schelgel et al.(1998)]{Sch98} Schlegel, D. J., Finkbeiner, D. P., Davis, M. 1998, ApJ, 500, 525
\bibitem[Skrutskie et al.(2006)] {skrut06} Skrutskie, M. F., Cutri, R. M., Stiening, R., et al. 2006, AJ, 131, 1163
\bibitem[Tuvikene et al.(2014)]{Tuv14} Tuvikene T., Edelmann, H., Groote, D., et al. 2014, Workflow for plate digitization, data extraction and publication, in Astroplate 2014, ed. L. Mi{\v s}kov\'a \& S. V\'itek (Institute of Chemical Technology, Prague), 2014, 127.
\bibitem[Vogt et al.(2016)]{vogt16} Vogt, N., Contreras-Quijada, A., Fuentes-Morales, I., et al. 2016, ApJS, 227, 6 
\bibitem[Watson et al.(2016)]{watson16} Watson, C., Henden, A. A., Price, A. 2016, AAVSO International Variable Star Index VSX, yCat, 102027
\bibitem[Whitelock et al.(1994)]{Whi94}Whitelock, P. A., Menzies, J., Feast, M. W., 1994, MNRAS 267, 711 
\bibitem[Whitelock et al.(2000)]{wmf00} Whitelock, P. A., Marang, F., Feast, M. W. 2000, MNRAS, 319, 728
\bibitem[Whitelock et al.(2003)]{Whi03} Whitelock, P.A., Feast, M.W., van Loon, J.Th., Zijlstra, A.A., 2003, MNRAS 342, 86
\bibitem[Whitelock(2012)] {white12} Whitelock, P. A. 2012, Ap\&SS, 341, 123
\bibitem[Wozniak et al.(2004a)]{woz04a} Wozniak, P. R., Vestrand, W. T., Akerlof, C. W., et al. 2004a, AJ, 127, 2436, http://skydot.lanl.gov/nsvs/nsvs.pht/
\bibitem[Wozniak et al.(2004b)]{woz04b} Wozniak, P. R., Williams, S. J., Vestrand, W. T., Gupta, V. 2004b, AJ, 128, 2965, VizieR On-line Data Catalog: J/AJ/128/2965
\bibitem[Wright et al.(2009)]{wright2009} Wright, N. J., Barlow, M. J., Drew, J. E., et al. 2009, MNRAS, 400, 1413
\end{thebibliography}
\end{document}